\documentclass[cameraready]{Interspeech}

\usepackage{makecell}
\usepackage{multirow}
\usepackage{booktabs}
\usepackage{tikz}
\usepackage{graphicx}
\usepackage{subcaption}
\usetikzlibrary{arrows.meta, positioning, fit, calc}
\newcommand{\mr}[1]{\multirow{2}{*}[-0.5ex]{#1}}

\AtBeginDocument{
  \setlength{\abovedisplayskip}{5pt}
  \setlength{\belowdisplayskip}{5pt}
  \setlength{\abovedisplayshortskip}{4pt}
  \setlength{\belowdisplayshortskip}{4pt}
}
\title{Neuromorphic Speech Enhancement with Dual-Branch Spiking Neural Networks}

\author[]{Taiyu}{Meng}
\author[correspondingauthor]{Wenbin}{Jiang}
\author[]{Haoyi}{Zhang}
\author[]{Yuhan}{Zhou}
\author[]{Haibing}{Yin}

\address{
    School of Communication Engineering, Hangzhou Dianzi University, Hangzhou, China
}

\email{\{meng\_taiyu, wbjiang, hyzhang11, yhzhou05, yhb\}@hdu.edu.cn}


\begin{document}
\maketitle

%======================================================================
\begin{abstract}
Spiking neural network (SNN)-based neuromorphic speech enhancement has emerged as a promising paradigm due to its energy efficiency, yet it still underperforms classical artificial neural network (ANN)-based approaches owing to binary activations and the lack of well-designed network architectures. To overcome this limitation, we propose a novel dual-branch spiking neural network architecture equipped with a gated spiking unit (GSU), termed GSU-DBNet. Specifically, GSU-DBNet simultaneously models the speech magnitude spectrum and complex spectrum, predicting the corresponding magnitude and complex spectral masks. Meanwhile, a dual-path GSU module is adopted to exploit temporal and frequency information for enhanced spatiotemporal feature representation. Experiments on a popular benchmark dataset show that GSU-DBNet achieves a PESQ score of 3.04 with only 394K parameters, outperforming existing SNN-based methods while using only 4.5\%--10.6\% of the parameters of representative ANN-based models.
\end{abstract}

%======================================================================
\section{Introduction}
\label{sec:intro}

Speech enhancement (SE) aims to recover clean speech from noisy observations and serves as a critical front-end for hearing aids, automatic speech recognition, and real-time communication. Deep neural networks have become the dominant paradigm~\cite{wang2018supervised,valentini2016investigating,jiang_unse_taslp}, achieving remarkable speech enhancement performance~\cite{zhang2026hyflowse,jiang2023}. However, current high-performance models typically require millions of parameters and intensive floating-point operations~\cite{hu2020dccrn,li2022gagnet,lu2023mpsenet}, whose computational and energy overhead is difficult to accommodate within the low-power, low-latency constraints of edge devices.

Spiking neural networks (SNNs) offer a fundamentally different paradigm: binary spikes and event-driven dynamics enable temporal modeling with fewer parameters and are natively compatible with low-power neuromorphic hardware~\cite{davies2018loihi}. With surrogate gradient training~\cite{neftci2019surrogate}, SNNs have achieved near-ANN performance on speech recognition~\cite{wu2020deep}, keyword spotting~\cite{bittar2022surrogate}, and other tasks, with recent extensions to speech enhancement~\cite{du2024spikings4}. The Gated Spiking Unit (GSU)~\cite{hao2025spikingfullsubnet} retains only a forget gate, halving the number of recurrent layer parameters relative to Long Short-Term Memory (LSTM) and providing a foundation for parameter-efficient SNN-based SE models. However, existing SNN-based SE methods generally adopt single branch architectures that model only a single spectral dimension. For example, Spiking-FullSubNet~\cite{hao2025spikingfullsubnet} focuses on complex spectral features and achieves parameter compression, but a notable quality gap with conventional networks remains. DPSNN~\cite{sun2024dpsnn} constructs a dual-path structure on time-domain waveforms and uniformly applies SNN units without investigating the suitability of spiking dynamics across signal domains. These methods fail to exploit the complementary advantages of magnitude and complex spectra in noise suppression and phase recovery.

In the conventional neural network domain, lightweight speech enhancement has developed effective architectural paradigms. Dual-path recurrent networks~\cite{luo2020dualpath,luo2019convtasnet} factorize long sequences into time and frequency dimensions for separate modeling, substantially reducing computational complexity. For spectral representation, methods such as complex spectral modeling~\cite{hu2020dccrn}, sub-band decomposition, dual-branch estimation~\cite{yu2022dualnet}, and multi-domain fusion~\cite{li2022gagnet,lu2023mpsenet} exploit complementary information across different representations to improve performance. However, these gains mainly stem from macro-level architectural changes, while the core recurrent modules remain largely unchanged. LSTM~\cite{hochreiter1997lstm} is still widely used, and even GRU~\cite{cho2014gru}, despite fewer gates, relies on continuous hidden states and dense floating-point operations, limiting energy efficiency. Integrating spiking units into these architectures could achieve both efficiency and performance gains, yet no prior work has systematically explored SNN modeling across feature dimensions in a dual-path framework.

Inspired by dual-path architectures~\cite{luo2020dualpath,jiang2025mos} and gated spiking modeling~\cite{hao2025spikingfullsubnet}, we propose \textbf{GSU-DBNet}, a dual-branch encoder--separator--decoder architecture for speech enhancement based on gated spiking units. Specifically, the separator employs a bidirectional BiGSU in the frequency path to capture global spectral correlations and a unidirectional GSU in the time path for causal temporal modeling. A dual-branch decoder simultaneously estimates a complex mask for phase-aware reconstruction via DeepFilter and a magnitude mask for energy envelope estimation, with the two outputs fused through weighted averaging. On VoiceBank+DEMAND, GSU-DBNet achieves PESQ~3.04 with only \textbf{394K parameters}, a parameter count of only 4.5\%--10.6\% of representative ANN methods, while improving PESQ by 0.84 and 0.38 over DPSNN~\cite{sun2024dpsnn} and Spiking-FSN~\cite{hao2025spikingfullsubnet} respectively. 

Overall, the main contributions of this work are threefold:
\begin{itemize}
  \item We propose GSU-DBNet, a dual-branch SNN architecture with gated spiking units, for joint magnitude–complex spectrum modeling and spatiotemporal feature extraction.
  \item GSU-DBNet achieves competitive speech quality with fewer parameters than ANN-based methods and demonstrates significant improvements over existing SNN approaches.
  \item Ablation studies reveal that the binary output bottleneck makes the single-gate design optimal, while additional gates degrade performance.
\end{itemize}

%======================================================================
\section{Proposed method}
\label{sec:method}

\subsection{Architecture overview}
\label{sec:overview}

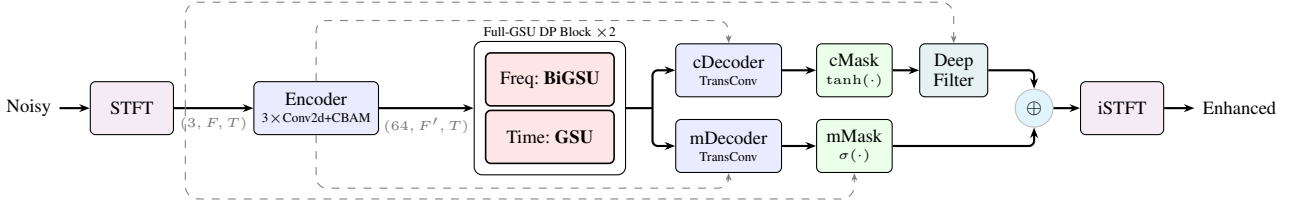
\begin{figure*}[tb]
\centering
\tikzstyle{sigblock} = [draw, rounded corners=2pt, minimum height=0.7cm,
    minimum width=1.1cm, font=\scriptsize, align=center]
\tikzstyle{dspblock} = [sigblock, fill=violet!8]
\tikzstyle{cnnblock} = [sigblock, fill=blue!8]
\tikzstyle{rnnblock} = [sigblock, fill=orange!12]
\tikzstyle{gsublock} = [sigblock, fill=red!10, line width=0.6pt]
\tikzstyle{maskblock}= [sigblock, fill=green!8]
\tikzstyle{dfblock}  = [sigblock, fill=teal!10]
\tikzstyle{iotext}   = [font=\scriptsize]
\tikzstyle{dimtext}  = [font=\tiny, text=black!55]
\tikzstyle{arr}      = [-{Stealth[length=4pt]}, thick]
\tikzstyle{skiparr}  = [-{Stealth[length=3pt]}, dashed, gray]
%--- (a) Architecture overview ---
\resizebox{\textwidth}{!}{%
\begin{tikzpicture}[node distance=0.45cm and 0.4cm]
% Input
\node[iotext] (noisy) {Noisy};
% STFT
\node[dspblock, right=0.4cm of noisy] (stft) {STFT};
% Encoder
\node[cnnblock, right=1.05cm of stft, minimum width=1.6cm] (enc)
    {Encoder\\[-1pt]{\tiny 3$\times$Conv2d+CBAM}};
% Inner freq/time paths (single-line horizontal layout)
\node[gsublock, right=1.4cm of enc, minimum width=1.7cm,
    yshift=0.38cm] (freq) {Freq:~\textbf{BiGSU}};
\node[gsublock, right=1.4cm of enc, minimum width=1.7cm,
    yshift=-0.38cm] (time) {Time:~\textbf{GSU}};
% Outer DP block frame
\node[draw, rounded corners=3pt, inner sep=4pt, fit=(freq)(time),
    label={[font=\tiny, yshift=-2pt]above:Full-GSU DP Block $\times$2}] (dpbox) {};
% Dual Decoders (two independent decoder paths)
\node[cnnblock, right=0.65cm of dpbox, minimum width=1.4cm,
    yshift=0.50cm] (dec1) {cDecoder\\[-1pt]{\tiny TransConv}};
\node[cnnblock, right=0.65cm of dpbox, minimum width=1.4cm,
    yshift=-0.50cm] (dec2) {mDecoder\\[-1pt]{\tiny TransConv}};
% Activation masks (function notation)
\node[maskblock, right=0.45cm of dec1, minimum width=1.0cm] (cmask)
    {cMask\\[-1pt]{\tiny $\tanh(\cdot)$}};
\node[maskblock, right=0.45cm of dec2, minimum width=1.0cm] (mmask)
    {mMask\\[-1pt]{\tiny $\sigma(\cdot)$}};
% DeepFilter (aligned with complex-mask path)
\node[dfblock, right=0.35cm of cmask, minimum width=0.9cm] (df)
    {Deep\\Filter};
% Merge node (⊕) at center level between decoder paths
\node[circle, draw=black!25, fill=cyan!10, minimum size=0.5cm,
    inner sep=0pt, font=\scriptsize, line width=0.4pt,
    right=0.35cm of df, yshift=-0.50cm] (merge) {$\oplus$};
% iSTFT
\node[dspblock, right=0.35cm of merge] (istft) {iSTFT};
% Enhanced
\node[iotext, right=0.4cm of istft] (enh) {Enhanced};
%--- Arrows (main flow) with dimension labels ---
\draw[arr] (noisy) -- (stft);
\draw[arr] (stft) -- (enc) node[dimtext, midway, below] {$(3,F,T)$};
\draw[arr] (enc) -- (dpbox) node[dimtext, midway, below] {$(64,F',T)$};
% DP block → two decoders (fan out)
\coordinate (forkpt) at ($(dpbox.east)+(0.35,0)$);
\draw[thick] (dpbox.east) -- (forkpt);
\draw[arr, rounded corners=2pt] (forkpt) |- (dec1.west);
\draw[arr, rounded corners=2pt] (forkpt) |- (dec2.west);
% Decoders → masks
\draw[arr] (dec1) -- (cmask);
\draw[arr] (dec2) -- (mmask);
% Complex path: cMask → DeepFilter → Merge
\draw[arr] (cmask) -- (df);
\draw[arr, rounded corners=2pt] (df.east) -| (merge);
% Magnitude path: mMask → Merge (bypasses DeepFilter)
\draw[arr, rounded corners=2pt] (mmask.east) -| (merge);
% Merge → iSTFT → Enhanced
\draw[arr] (merge) -- (istft);
\draw[arr] (istft) -- (enh);
% Skip connections (encoder → each decoder)
\draw[skiparr, rounded corners=4pt] (enc.north) -- ++(0,0.8) -| (dec1.north);
\draw[skiparr, rounded corners=4pt] (enc.south) -- ++(0,-0.7) -| (dec2.south);
% Noisy-signal bypass to DeepFilter (for coefficient-based filtering)
\draw[skiparr, rounded corners=5pt]
    ($(stft.east)+(0.15,0)$) -- ++(0,1.4) -| (df.north);

% Noisy-signal bypass to mMask (for magnitude masking)
\draw[skiparr, rounded corners=5pt]
    ($(stft.east)+(0.15,0)$) -- ++(0,-1.2) -| (mmask.south);
\end{tikzpicture}%
}% end resizebox

\caption{Overview of the GSU-DBNet architecture with dual-path spiking blocks and a complex-magnitude dual-branch decoder.}
\label{fig:architecture}
\end{figure*}

As illustrated in Figure~\ref{fig:architecture}, GSU-DBNet follows an encoder--separator--decoder paradigm. The noisy speech is first transformed via STFT, from which the real part, imaginary part, and magnitude spectrum are extracted and concatenated into a three-channel spectral input. The encoder comprises three convolutional blocks, each consisting of a Conv2d layer, GroupNorm, PReLU, and a CBAM~\cite{woo2018cbam} attention module. The first two layers progressively compress the frequency dimension and increase the channel count through strided convolutions, while the third layer employs a $1{\times}1$ convolution to increase the channel count to 64, producing a latent feature map. Subsequently, two stacked dual-path GSU blocks alternately process the encoded features along the frequency and time dimensions. The frequency path uses BiGSU to model cross-frequency dependencies, whereas the time path uses GSU to capture causal temporal dependencies. Each path is followed by a linear projection layer, GroupNorm, and a residual connection. Finally, two independent transposed-convolution decoders incorporate U-Net~\cite{ronneberger2015unet} skip connections to fuse features from the corresponding encoder layers. The complex decoder applies a tanh activation function to generate DeepFilter~\cite{schroeter2022deepfilternet} coefficients for filtering the noisy STFT, while the magnitude decoder applies a sigmoid activation function to generate a magnitude mask that is multiplied with the noisy magnitude spectrum. The outputs of the two branches are fused via weighted averaging, and the enhanced speech is reconstructed through the inverse STFT.

\subsection{Dual-branch spectral processing}
\label{sec:dualbranch}

GSU-DBNet employs a dual-branch decoding structure that performs parallel modeling of the complex and magnitude spectra, fully exploiting the complementary information provided by these two spectral representations.

The complex branch jointly estimates the real and imaginary components of the spectrum, thereby preserving phase information. This branch progressively recovers frequency resolution through a transposed-convolution decoder and fuses features from the corresponding encoder layers via U-Net skip connections. The decoder output is passed through a tanh activation function to generate a complex mask, which serves as DeepFilter~\cite{schroeter2022deepfilternet} coefficients for filtering the noisy STFT spectrum. The magnitude branch focuses on accurately estimating the speech energy envelope, adopting the same decoder and skip-connection structure with a sigmoid activation function to generate a magnitude mask. The two branches produce spectral estimates as follows:
\begin{equation}
    \begin{aligned}
        Y_c &= \text{DeepFilter}\bigl(\tanh(D_c(Z)), X\bigr), \\
        Y_m &= \sigma\bigl(D_m(Z)\bigr) \odot |X|,
    \end{aligned}
    \label{eq:dual}
\end{equation}
where $Z$ denotes the separator output and $X$ the noisy STFT. The two estimates are fused via weighted averaging and converted to the enhanced waveform through inverse STFT:
\begin{equation}
  \label{eq:fusion}
  Y = \alpha Y_c + (1-\alpha) Y_m.
\end{equation}
The complex branch excels at correcting phase information, while the magnitude branch provides stable energy estimation, and their fusion enables the model to benefit from the complementary strengths of both branches.

\subsection{Gated spiking unit}
\label{sec:gsu}
We adopt the gated spiking unit~\cite{hao2025spikingfullsubnet} as the fundamental recurrent cell in the dual-path module. GSU is a recurrent unit inspired by the leaky integrate-and-fire (LIF) neuron~\cite{burkitt2006lif}, which maintains a membrane potential through a gated memory mechanism and transmits information via binary spikes. Its computation involves two key mechanisms: gated membrane potential updates and spike emission.

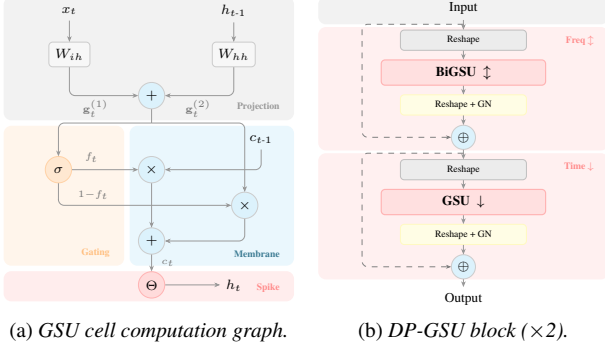
\begin{figure}[t]
\centering
% --- (a) GSU cell computation graph ---
\begin{subfigure}[t]{0.48\columnwidth}
\centering
\resizebox{\linewidth}{!}{%
\begin{tikzpicture}[
    op/.style={circle, draw=black!20, minimum size=0.48cm, inner sep=0pt,
        font=\scriptsize, line width=0.4pt},
    addop/.style={op, fill=cyan!12},
    mulop/.style={op, fill=cyan!12},
    gatop/.style={op, fill=orange!18, draw=orange!35},
    spikeop/.style={op, fill=red!14, draw=red!30, line width=0.5pt},
    wt/.style={draw=black!20, rounded corners=2pt, minimum height=0.45cm,
        minimum width=0.75cm, font=\scriptsize, fill=gray!6, line width=0.4pt},
    io/.style={font=\scriptsize},
    note/.style={font=\tiny, text=black!55},
    conn/.style={-{Stealth[length=2.8pt, width=2.2pt]}, black!40, thin},
    pln/.style={black!40, thin},
    grplabel/.style={font=\tiny\bfseries},
]
\useasboundingbox (-0.2,-0.75) rectangle (5.1,4.85);
% ======== Background regions ========
\fill[gray!10, rounded corners=4pt] (-0.2,2.6) rectangle (5.1,4.85);
\fill[orange!8, rounded corners=4pt] (-0.2,-0.05) rectangle (2.0,2.5);
\fill[cyan!7, rounded corners=4pt] (2.1,-0.05) rectangle (5.1,2.5);
\fill[red!7, rounded corners=4pt] (-0.2,-0.7) rectangle (5.1,-0.15);
% ======== Region labels ========
\node[grplabel, text=black!40, anchor=south east] at (4.95,2.65) {Projection};
\node[grplabel, text=orange!50, anchor=south east] at (1.85,0.00) {Gating};
\node[grplabel, text=cyan!50!black, anchor=south east] at (4.95,0.00) {Membrane};
\node[grplabel, text=red!45, anchor=south east] at (4.95,-0.65) {Spike};
%
% ======== Nodes ========
\node[io] (xt) at (1.0,4.6) {$x_t$};
\node[io] (ht1) at (4.0,4.6) {$h_{t\text{-}1}$};
\node[wt] (wih) at (1.0,3.8) {$W_{ih}$};
\node[wt] (whh) at (4.0,3.8) {$W_{hh}$};
\node[addop] (add1) at (2.5,3.05) {$+$};
\node[gatop] (sigma) at (0.8,1.7) {$\sigma$};
\node[io] (ct1) at (4.5,2.25) {$c_{t\text{-}1}$};
\node[mulop] (mul1) at (2.5,1.7) {$\times$};
\node[mulop] (mul2) at (4.2,1.05) {$\times$};
\node[addop] (add2) at (2.5,0.4) {$+$};
\node[spikeop] (theta) at (2.5,-0.4) {$\Theta$};
\node[io] (ht) at (4.0,-0.4) {$h_t$};
%
% ======== Arrows ========
\draw[conn] (xt) -- (wih);
\draw[conn] (ht1) -- (whh);
\draw[conn, rounded corners=3pt] (wih.south) |- (add1.west);
\draw[conn, rounded corners=3pt] (whh.south) |- (add1.east);
\coordinate (sp) at (2.5,2.55);
\draw[pln] (add1.south) -- (sp);
% Split to gate and cell input
\draw[conn, rounded corners=3pt] (sp) -| (sigma.north)
    node[note, pos=0.3, above] {$\mathbf{g}_t^{(1)}$};
\draw[conn, rounded corners=3pt] (sp) -| (mul2.north)
    node[note, pos=0.25, above] {$\mathbf{g}_t^{(2)}$};
% Gate outputs
\draw[conn] (sigma.east) -- (mul1.west)
    node[note, midway, above, xshift=-7pt, yshift=1pt] {$f_t$};
\draw[conn, rounded corners=3pt] (ct1.south) |- (mul1);
\draw[conn, rounded corners=3pt] (sigma.south) -- ++(0,-0.20) |- (mul2.west)
    node[note, pos=0.20, left, xshift=29pt, yshift=2pt] {$1{-}f_t$};
% Membrane computation
\draw[conn] (mul1.south) -- (add2.north);
\draw[conn, rounded corners=3pt] (mul2.south) |- (add2.east);
\draw[conn] (add2.south) -- (theta.north)
    node[note, midway, right, xshift=2pt] {$c_t$};
% Spike output
\draw[conn] (theta.east) -- (ht);
\end{tikzpicture}%
}%
\caption{GSU cell computation graph.}
\label{fig:gsu-cell}
\end{subfigure}
\hfill
% --- (b) DP-GSU block (vertical flow) ---
\begin{subfigure}[t]{0.48\columnwidth}
\centering
\resizebox{\linewidth}{!}{%
\begin{tikzpicture}[
    op/.style={circle, draw=black!20, minimum size=0.44cm, inner sep=0pt,
        font=\scriptsize, line width=0.4pt},
    addop/.style={op, fill=cyan!12},
    block/.style={draw=black!25, rounded corners=2pt,
        minimum height=0.38cm, font=\scriptsize, align=center, line width=0.4pt},
    freqb/.style={block, fill=blue!20, draw=blue!35, minimum width=3.0cm,
        minimum height=0.46cm},
    gsublock/.style={block, fill=red!12, draw=red!30, minimum width=3.0cm,
        minimum height=0.46cm, line width=0.6pt},
    reshb/.style={block, fill=gray!15, minimum width=2.3cm,
        minimum height=0.28cm, font=\tiny},
    utilb/.style={block, fill=yellow!12, draw=yellow!30, minimum width=2.3cm,
        minimum height=0.28cm, font=\tiny},
    arr/.style={-{Stealth[length=2.8pt, width=2.2pt]}, black!40, thin},
    skiparr/.style={-{Stealth[length=2.5pt]}, dashed, black!50, thin},
    io/.style={font=\scriptsize},
    grplabel/.style={font=\tiny\bfseries},
]
\useasboundingbox (-0.2,-0.75) rectangle (5.1,4.85);
% ======== Background regions ========
\fill[gray!10, rounded corners=4pt] (-0.2,4.35) rectangle (5.1,4.85);
\fill[red!6, rounded corners=4pt] (-0.2,2.05) rectangle (5.1,4.30);
\fill[red!6, rounded corners=4pt] (-0.2,-0.30) rectangle (5.1,2.00);
%
% ======== Region labels ========
\node[grplabel, text=red!45, anchor=north east] at (4.95,4.25) {Freq\,$\updownarrow$};
\node[grplabel, text=red!45, anchor=north east] at (4.95,1.95) {Time\,$\downarrow$};
%
% ======== Nodes ========
\node[io] (input) at (2.45,4.65) {Input};
\node[reshb] (resh1) at (2.45,4.06) {Reshape};
% --- Freq path ---
\node[gsublock] (bilstm) at (2.45,3.47) {\textbf{BiGSU}\enspace$\updownarrow$};
\node[utilb] (pn1) at (2.45,2.88) {Reshape + GN};
\node[addop] (add1) at (2.45,2.29) {$\oplus$};
% --- Time path ---
\node[reshb] (resh3) at (2.45,1.70) {Reshape};
\node[gsublock] (gsu) at (2.45,1.11) {\textbf{GSU}\enspace$\downarrow$};
\node[utilb] (pn2) at (2.45,0.52) {Reshape + GN};
\node[addop] (add2) at (2.45,-0.07) {$\oplus$};
% --- Output ---
\node[io] (out) at (2.45,-0.66) {Output};
%
% ======== Arrows ========
\draw[arr] (input) -- (resh1);
\draw[arr] (resh1) -- (bilstm);
\draw[arr] (bilstm) -- (pn1);
\draw[arr] (pn1) -- (add1);
\draw[arr] (add1) -- (resh3);
\draw[arr] (resh3) -- (gsu);
\draw[arr] (gsu) -- (pn2);
\draw[arr] (pn2) -- (add2);
\draw[arr] (add2) -- (out);
% ======== Residual connections (from 4D wire, before reshape) ========
\coordinate (fork1) at (2.45,4.35);   % on Input→Reshape wire (4D)
\coordinate (fork2) at (2.45,2.00);   % on ⊕₁→Reshape wire (4D)
\draw[skiparr, rounded corners=3pt]
    (fork1) -- ++(-1.85,0) |- (add1.west);
\draw[skiparr, rounded corners=3pt]
    (fork2) -- ++(-1.85,0) |- (add2.west);
\end{tikzpicture}%
}%
\caption{DP-GSU block ($\times$2).}
\label{fig:dpblock}
\end{subfigure}

\caption{(a)~GSU cell computation graph. (b)~DP-GSU block with BiGSU (frequency) and GSU (time).}
\label{fig:gsu}
\vspace{-10pt}
\end{figure}

As illustrated in Figure~\ref{fig:gsu}(a), the GSU updates its membrane potential through a single-gate mechanism. At each time step $t$, given the input $x_t$ and the previous spike output $h_{t-1}$, the GSU computes a joint linear projection and splits it into two halves. A forget gate $f_t$ controls the temporal decay of the membrane potential, where values close to 1 preserve the previous state and values close to 0 favor the incorporation of new input. The complementary term $1-f_t$ serves as an implicit input gate, thereby reducing the number of recurrent-layer parameters by approximately half compared with an LSTM. The projection and membrane potential update are given by:
\begin{equation}
    \begin{aligned}
        \mathbf{g}_t &= W_{ih}\, x_t + W_{hh}\, h_{t-1} + b, \\
        f_t &= \sigma\bigl(\mathbf{g}_t^{(1)}\bigr), \\
        c_t &= f_t \odot c_{t-1} + (1 - f_t) \odot \mathbf{g}_t^{(2)}.
    \end{aligned}
    \label{eq:snn_gate}
\end{equation}

The GSU transmits information through binary spikes. The membrane potential $c_t$ is passed through the Heaviside step function $\Theta$ to produce the binary output $h_t$. Since $\Theta$ is non-differentiable, a triangular surrogate gradient~\cite{neftci2019surrogate} is employed during training for approximation:
\begin{equation}
\begin{aligned}
    h_t = \Theta(c_t), \quad \frac{\partial \Theta}{\partial c_t} \approx \frac{1}{\gamma^2}\max\bigl(\gamma - \lvert c_t\rvert, 0\bigr),
\end{aligned}
\label{eq:spike_surrogate}
\end{equation}
with $\gamma = 1.0$, enabling standard backpropagation through time~\cite{werbos1990bptt}. Each neuron transmits only one bit of information per time step. This binary output bottleneck is the fundamental source of GSU's parameter efficiency, while also limiting the benefits of additional gates.

% To verify this, we define two multi-gate spiking variants inspired by spiking LSTM~\cite{lotfi2020spikinglstm} for ablation. SLSTM-2G decouples the forget and input gates into two independent sigmoid gates, with the projection dimension expanded to $3H$. SLSTM-3G further adds an output gate that weights the membrane potential before thresholding, with the projection dimension expanded to $4H$. All three variants share the forget gate and spike output mechanism, differing only in the number of gates and membrane potential computation:
To verify this hypothesis, we define two multi-gate spiking variants inspired by spiking LSTM~\cite{lotfi2020spikinglstm} for ablation. SLSTM-2G decouples the forget and input gates into two independent sigmoid gates, with the projection dimension expanded to $3H$. SLSTM-3G further introduces an output gate that modulates the membrane potential before thresholding, with the projection dimension expanded to $4H$. All three variants share the same forget-gate and spike-generation mechanisms, differing only in the number of gates and the membrane potential update:
\begin{equation}
  i_t = \sigma(\mathbf{g}_t^{(2)}),\,
  c_t = f_t \odot c_{t-1} + i_t \odot \mathbf{g}_t^{(3)},\,
  h_t = \Theta(c_t).
  \label{eq:slstm2g}
\end{equation}
\begin{equation}
  o_t = \sigma(\mathbf{g}_t^{(3)}),\,
  c_t = f_t \odot c_{t-1} + i_t \odot \mathbf{g}_t^{(4)},\,
  h_t = \Theta(o_t {\odot} c_t).
  \label{eq:slstm3g}
\end{equation}

\subsection{Loss function}
\label{sec:loss}

The model is trained using a hybrid loss function that combines power-law-compressed spectral MSE and time-domain SI-SNR~\cite{leroux2019sisnr}. The spectral loss applies power-law compression to the real part, imaginary part, and magnitude spectrum of the STFT before computing the MSE. Denoting the magnitude by $M$, the phase by $\theta$, and the compression exponent by $c{=}0.3$, the compressed representations are defined as:
\begin{equation}
\tilde{r} = M^c\cos\theta, \quad
\tilde{i} = M^c\sin\theta, \quad
\tilde{m} = M^c.
\label{eq:compress}
\end{equation}
This compression emphasizes low-energy spectral regions, thereby improving perceptual quality. The time-domain loss is based on the scale-invariant signal-to-noise ratio (SI-SNR), computed from the waveform reconstructed via the inverse STFT. The total loss is given by
\begin{equation}
\mathcal{L} = \alpha_c\bigl(\lVert\Delta\tilde{r}\rVert^2 {+} \lVert\Delta\tilde{i}\rVert^2\bigr) + \alpha_m\lVert\Delta\tilde{m}\rVert^2 + \mathcal{L}_{\text{SI-SNR}},
\label{eq:loss}
\end{equation}
where $\alpha_c{=}30$ and $\alpha_m{=}70$ are determined through preliminary experiments and fixed across all configurations.

%======================================================================
\section{Experiments}
\label{sec:experiments}
%======================================================================

\subsection{Dataset}

We evaluate our method on the VoiceBank+DEMAND dataset~\cite{valentini2016investigating}, which comprises 11,572 training utterances from 28 speakers mixed with 10 noise types at SNRs of 0, 5, 10, and 15 dB. The test set contains 824 utterances from two held-out speakers mixed with five unseen noise types at SNRs of 2.5, 7.5, 12.5, and 17.5 dB, all sampled at 16 kHz.

\subsection{Implementation details}

The STFT uses a Hann window of 400 samples (\SI{25}{\milli\second}), a hop size of 100 samples (\SI{6.25}{\milli\second}), and an FFT size of 512, producing a 257-bin spectrum.

The encoder consists of three convolutional blocks, each comprising a Conv2d layer, GroupNorm, PReLU, and a CBAM attention module, with output channels of 16, 32, and 64, respectively. The kernel sizes and strides of the first two layers are $(5,2)$ and $(2,1)$, respectively, while the third layer uses a kernel size and stride of $(1,1)$. The decoder is symmetric to the encoder, employing transposed convolutions to progressively recover the frequency resolution, with U-Net skip connections used to fuse features from the corresponding encoder layers. The separator contains two stacked dual-path GSU blocks, with the hidden dimensions of both the frequency and time paths set to 128. The output is passed through a PReLU layer followed by a $1{\times}1$ convolution. The frequency, time, and channel filter orders of DeepFilter are set to 3, 5, and 16, respectively.

The model is trained using AdamW~\cite{loshchilov2019adamw} with an initial learning rate of $1\times10^{-3}$, which is reduced by a factor of 0.5 using ReduceLROnPlateau (patience = 10). Gradient norms are clipped to 5.0. The batch size is 18, and training is performed for up to 150 epochs.

We evaluate the models using wideband PESQ~\cite{rix2001pesq}, the composite measures CSIG, CBAK, and COVL~\cite{hu2008evaluation}, as well as segmental SNR (SSNR). The ablation experiments additionally report DNSMOS~\cite{reddy2022dnsmos} SIG, BAK, and OVRL scores, STOI~\cite{taal2011stoi} (\%), and SI-SNR (dB). Audio samples from the compared methods are available online\footnote{\url{https://meng-taiyu.github.io/dpnet-demo/}}.

%======================================================================
\section{Results}
\label{sec:results}
%======================================================================

\subsection{Comparison with existing methods}

\begin{table}[t]
  \caption{Comparison on VoiceBank+DEMAND. $^\dagger$Published results. $^\ddagger$Reproduced. Best per metric in \textbf{bold}.}
  \label{tab:main}
  \setlength{\tabcolsep}{3.0pt}
  \renewcommand\arraystretch{1.1}
  \centering
  \resizebox{\columnwidth}{!}{%
  \begin{tabular}{@{}lcccccc@{}}
    \toprule
    \textbf{Method} & \textbf{\makecell{\#Params. \\(K)}} & \textbf{PESQ} & \textbf{CSIG} & \textbf{CBAK} & \textbf{COVL} & \textbf{SSNR} \\
    \midrule
    Noisy & -- & 1.97 & 3.35 & 2.44 & 2.63 & 1.68 \\
    \midrule
    DCCRN$^\dagger$~\cite{hu2020dccrn}           & 3700 & 2.68 & 3.88 & 3.18 & 3.27 & 8.62 \\
    FullSubNet+$^\dagger$~\cite{chen2022fullsubnetplus} & 8670 & 2.88 & 3.86 & 3.42 & 3.57 & -- \\
    GaGNet$^\dagger$~\cite{li2022gagnet}         & 5940 & 2.94 & 4.26 & 3.45 & 3.59 & -- \\
    TSTNN$^\dagger$~\cite{wang2021tstnn}         & 920  & 2.96 & \textbf{4.33} & 3.53 & 3.67 & 9.70 \\
    DPSNN$^\ddagger$~\cite{sun2024dpsnn}          & 572  & 2.20 & 3.21 & 2.99 & 2.68 & 8.30 \\
    Spiking-FSN$^\ddagger$~\cite{hao2025spikingfullsubnet} & 954 & 2.66 & 3.85 & 3.24 & 3.24 & 8.31 \\
    \midrule
    \textbf{GSU-DBNet (ours)} & \textbf{394} & \textbf{3.04} & 4.28 & \textbf{3.57} & \textbf{3.68} & \textbf{9.94} \\
    \bottomrule
  \end{tabular}}%
  \vspace{-10pt}
\end{table}

Table~\ref{tab:main} summarizes a comparison between GSU-DBNet and representative ANN-based methods and SNN baselines. GSU-DBNet achieves a PESQ score of 3.04 with only 394K parameters and attains the best scores on CBAK, COVL, and SSNR, indicating that dual-branch spectral modeling provides consistent improvements in both perceptual quality and noise suppression. 

Compared with larger-scale ANN-based methods, GSU-DBNet surpasses DCCRN (2.68), FullSubNet+ (2.88), and GaGNet (2.94) in terms of PESQ while requiring 9--22$\times$ fewer parameters. Compared with TSTNN, which also operates with fewer than one million parameters, GSU-DBNet improves the PESQ score from 2.96 to 3.04 while using fewer parameters, achieving better overall enhancement quality. However, TSTNN retains a slight advantage in CSIG, suggesting that there is still room for improvement in speech signal consistency.

Compared with SNN baselines, GSU-DBNet improves PESQ by 0.84 over DPSNN and by 0.38 over Spiking-FSN, demonstrating that dual-path gated spiking modeling effectively narrows the performance gap with mainstream methods while maintaining a low parameter overhead. Overall, GSU-DBNet achieves the best results on most evaluation metrics while using the fewest parameters among all compared methods.

\subsection{Ablation study}
\label{sec:ablation}

\begin{table}[!t]
  \caption{Ablation study on VoiceBank+DEMAND. The baseline is the same GSU-DBNet as in Table~\ref{tab:main}. Best per column in \textbf{bold}.}
  \label{tab:ablation}
  \setlength{\tabcolsep}{2.0pt}
  \renewcommand\arraystretch{1.1}
  \centering
  \resizebox{\columnwidth}{!}{%
  \begin{tabular}{@{}lccccccc@{}}
    \toprule
    \mr{\textbf{Config}} & \mr{\textbf{\makecell{\#Params. \\ (K)}}} & \mr{\textbf{PESQ}} & \multicolumn{3}{c}{\textbf{DNSMOS}} & \mr{\textbf{\makecell{STOI\\(\%)}}} & \mr{\textbf{\makecell{SI-SNR \\ (dB)}}} \\
    \cmidrule(lr){4-6}
    &  & & \textbf{SIG} & \textbf{BAK} & \textbf{OVRL} &  & \\
    \midrule
    GSU-DBNet              & 394 & 3.04 & \textbf{3.44} & 4.02 & \textbf{3.15} & 94.8 & 18.87 \\
    \midrule
    \multicolumn{8}{l}{\textit{(a) Branch ablation}} \\
    \quad w/o Mag-branch   & 379 & 2.96 & 3.42 & \textbf{4.03} & \textbf{3.15} & 94.6 & \textbf{19.02} \\
    \quad w/o Comp-branch  & 379 & 2.94 & 3.39 & 4.01 & 3.11 & 94.4 & 18.16 \\
    \midrule
    \multicolumn{8}{l}{\textit{(b) Path ablation}} \\
    \quad w/o Time-path    & 278 & 2.80 & 3.40 & 3.98 & 3.09 & 94.2 & 18.90 \\
    \quad w/o Freq-path    & 163 & 2.72 & 3.38 & 3.95 & 3.07 & 93.3 & 18.77 \\
    \midrule
    \multicolumn{8}{l}{\textit{(c) Cell design}} \\
    \quad Full-SLSTM2G     & 542 & 3.04 & \textbf{3.44} & 4.02 & \textbf{3.15} & 94.7 & 18.88 \\
    \quad Full-SLSTM3G     & 690 & 2.98 & 3.43 & 4.01 & 3.14 & 94.6 & 18.83 \\
    \midrule
    \multicolumn{8}{l}{\textit{(d) Model capacity}} \\
    \quad H=64             & 163 & 2.86 & 3.41 & 4.01 & 3.12 & 94.4 & 19.01 \\
    \quad H=192            & 714 & \textbf{3.07} & \textbf{3.44} & 4.02 & \textbf{3.15} & \textbf{94.9} & 18.98 \\
    \bottomrule
  \end{tabular}}%
  \vspace{-10pt}
\end{table}

We conduct ablation studies along four dimensions to evaluate the effectiveness of each design choice, with the results presented in Table~\ref{tab:ablation}.

\textit{(a)} We separately remove the magnitude branch (complex-only) and the complex branch (magnitude-only) to verify the necessity of dual-branch joint estimation. Both single-branch variants yield lower PESQ scores than the dual-branch baseline (2.96 and 2.94 vs.\ 3.04), confirming that each branch captures complementary spectral information. Removing the complex branch causes a larger degradation, with SI-SNR dropping by 0.71~dB, whereas SI-SNR remains nearly unchanged in the complex-only variant. This indicates that complex mask estimation plays a central role in waveform reconstruction through its phase-modeling capability.

\textit{(b)} We remove the time and frequency paths to study their roles in the dual-path structure. Both are essential for enhancement, as removing either significantly degrades PESQ, with the frequency path causing a larger drop. This is expected, since the frequency path captures global spectral structure via bidirectional modeling, contributing more to noise suppression and harmonic recovery, while the time path enforces inter-frame continuity through causal modeling. Together, they form a complete time--frequency modeling framework.

\textit{(c)} We replace GSU in both paths with multi-gate spiking variants SLSTM-2G and SLSTM-3G to study the effect of gate complexity. As the number of gates increases from 1 to 3, parameters nearly double, yet PESQ does not improve and instead declines. The two-gate variant matches GSU, while the three-gate variant degrades noticeably. We attribute this to the binary output bottleneck: each neuron outputs only 1~bit per time step, and although additional gates refine membrane potential dynamics, this information is compressed into binary spikes via Heaviside thresholding and cannot be effectively propagated. This suggests that the simplest single-gate design is optimal within the spiking framework.

\textit{(d)} We vary the hidden dimension $H$. Doubling $H$ from a smaller value substantially improves PESQ, but further increases yield diminishing returns while significantly increasing the parameter count, consistent with the binary output bottleneck discussed in (c). We therefore select $H{=}128$, which lies near the inflection point, as the default configuration.

\begin{figure}[t]
\centering
\includegraphics[width=\columnwidth,trim=10 0 5 5,clip]{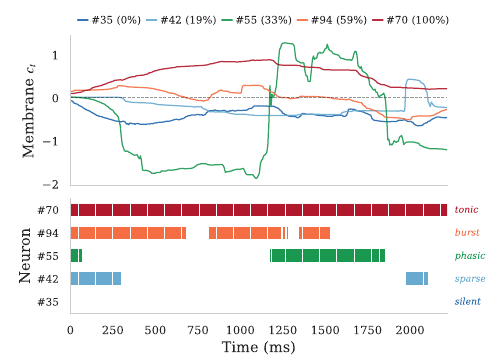}
\caption{Membrane potential and spike raster of the time-path GSU. Five representative neurons from 128 hidden units are selected, showing diverse self-organized firing patterns ranging from silent to tonic firing.}
\label{fig:spike}
\vspace{-15pt}
\end{figure}

\subsection{Spike activity analysis}
\label{sec:spike}

Figure~\ref{fig:spike} visualizes the membrane potential and spike raster of the first-layer time-path GSU on a test utterance. Five representative neurons out of 128 hidden units are selected, revealing that the network self-organizes into diverse firing patterns ranging from silent, sparse, phasic, burst, to tonic firing, with firing rates of 0\%, 19\%, 33\%, 59\%, and 100\%, respectively. Across all 128 neurons and 824 test utterances, the mean firing rate is 37\% with a standard deviation of 19\%, confirming that this diversity is systematic rather than incidental.

This observation confirms the ablation study in Section~\ref{sec:ablation}: although each neuron transmits only one bit per step, the single-gate GSU enables a rich functional spectrum through population coding, where 128 neurons with heterogeneous firing rates collectively express temporal features. Since the bottleneck lies in one-bit output thresholding rather than input gating precision, fine-grained modulation from additional gates is discarded after thresholding and thus provides no benefit. The 37\% mean firing rate further implies that two-thirds of neurons are silent at any time step, confirming that the network relies on sparse population coding rather than dense activation, a property well suited to event-driven neuromorphic hardware.

%======================================================================
 \section{Conclusions}
\label{sec:conclusion}

We propose GSU-DBNet, which integrates the Gated Spiking Unit into a dual-path, dual-branch architecture as a replacement for LSTM, enabling joint magnitude and complex mask estimation. Experiments on VoiceBank+DEMAND show that GSU-DBNet achieves a PESQ of 3.04 with only 394K parameters, improving by 0.84 and 0.38 over DPSNN and Spiking-FSN, respectively, while using only 4.5\%--10.6\% of representative ANN parameters. Furthermore, ablation studies show that the binary output bottleneck makes the single-gate design optimal, as additional gates do not improve performance but add redundant parameters, providing empirical evidence for spiking recurrent cell design. Future work includes conducting experiments on more diverse datasets and exploring deployment on neuromorphic hardware.

%======================================================================
\clearpage

\section{Acknowledgments}
This work was supported in part by Yangtze River Delta Science and Technology Innovation Community Joint Research under Grant 2024CSJGG1100, in part by the Zhejiang Provincial Natural Science Foundation of China (No. LMS26F020008), and in part by the Zhejiang Provincial College Student Innovation and Entrepreneurship Training Program under Grant S202510336076.

\section{Generative AI Use Disclosure}
Generative AI tools were used for editing and polishing the manuscript text. All experimental results, method design, and scientific conclusions are the sole responsibility of the authors.

\bibliographystyle{IEEEtran}
\bibliography{refs}

\end{document}